\documentclass[aps,nofootinbib,showpacs,preprint]{revtex4}
\def\dd{{\rm d}}

\def\_#1{^{}_{#1}}

\def\beq{\begin{equation}}
\def\eeq{\end{equation}}
\def\bea{\begin{eqnarray}}
\def\eea{\end{eqnarray}}

\def\dd{{\rm d}}

\def\_#1{^{}_{#1}}

\def\delp{{\Delta p}}
\def\delx{{\Delta x}}

\usepackage{enumerate}
\begin{document}

\title{(Anti-)de Sitter Black Hole Thermodynamics and the Generalized
Uncertainty Principle}
\author{Brett Bolen\footnote{E-mail: brett@phy.olemiss.edu},
Marco Cavagli\`a\footnote{E-mail: cavaglia@phy.olemiss.edu}}
\affiliation{Department of Physics and Astronomy,
University of Mississippi\\ University, MS 38677-1848, U.S.A.}

\begin{abstract}

We extend the derivation of the Hawking temperature of a Schwarzschild black
hole via the Heisenberg uncertainty principle to the de Sitter and anti-de
Sitter spacetimes. The thermodynamics of the Schwarzschild-(anti-)de Sitter
black holes is obtained from the generalized uncertainty principle of string
theory and non-commutative geometry. This may explain why the thermodynamics of
(anti-)de Sitter-like black holes admits a holographic description in terms of
a dual quantum conformal field theory, whereas the thermodynamics of
Schwarzschild-like black holes does not.

\end{abstract}

\maketitle

\section{Introduction}

The Heisenberg uncertainty principle of quantum mechanics allows a heuristic
derivation of the Hawking temperature \cite{Hawking-temp} of a Schwarzschild
black hole. The derivation proceeds as follows \cite{Adler:2001vs}. The
uncertainty in the linear position $x$ of an emitted quantum is approximately
equal to the Schwarzschild radius $r_s$. By modelling the black hole as an
object with linear size $r_s$, and assuming that the radiation satisfies the
condition of minimum uncertainty, the uncertainty in the energy of the emitted
quanta is
\beq
\Delta E\sim c\Delta p \sim \frac{\hbar c}{\Delta x}
\sim \frac{\hbar c}{r_s}\,,\qquad \to\qquad \Delta E=\kappa\frac{\hbar c}{r_s}\,,
\label{DE}
\eeq
where $\kappa$ is a proportionality constant. $\Delta E$ is identified with the
temperature $T$ of the radiation. Setting $\kappa=(d-3)/4\pi$, Eq.~(\ref{DE})
gives the Hawking temperature for a $d$-dimensional Schwarzschild black hole
\beq
T_H=\frac{d-3}{4\pi r_s}\hbar c\,.
\label{TS}
\eeq
The above derivation deserves some comments. Black hole emission is usually
regarded as being originated by quantum effects in the region around the black
hole horizon, such as semiclassical wave scattering or particle tunnelling.
(See, e.g.\ Ref.~\cite{Novikov:sz}.) The uncertainty principle does not
describe the origin of these effects, but only their consequence on the
measurement process. Explaining the origin of black hole emission requires the
knowledge of the quantum states that describe the black hole, from which the
exact form of the uncertainty principle for the black hole can be derived. On
the other hand, Eq.~(\ref{DE}) seems to suggest that black hole thermodynamics
is a generic low-energy effect of small scale physics. Since any quantum theory
of gravity must include some kind of uncertainty principle that reduces to
Heisenberg principle at low-energy scales, black hole thermodynamics should not
depend too much on the details of the quantum gravity theory. This seems to
agree in spirit with Visser's conclusion that the Hawking radiation only
requires ordinary quantum mechanics plus a slowly evolving future horizon, and
thus the knowledge of quantum gravity is unnecessary to explain the features of
black hole thermodynamics \cite{Visser:2001kq}. \\ 

The above derivation, although appealing, is only known for the Schwarzschild
black hole. The aim of paper is to extend the uncertainty principle derivation
of the Hawking temperature to the de-Sitter (dS) and anti-de Sitter (adS) black
holes.

\section{AdS and dS Thermodynamics}

The line element of a $d$-dimensional Schwarzschild-(a)dS black hole ($d>3$)
with mass $M$ is (see, e.g., Refs.~\cite{Cadoni:2003pp,Cai:2001sn})
\beq
ds^2 = - \left( 1 \pm \lambda^2  r^2 - \frac{\omega_d G_d M}{c^2r^{d-3}}
\right) c^2\dd t^2 +\left( 1 \pm \lambda^2  r^2 - \frac{\omega_d G_d M}{c^2r^{d-3}}
\right)^{-1} \dd  r^2 +  r^2 \dd \Omega^2_{d -2}\,,
\eeq
where $G_d$ is Newton's constant, $\lambda = 1/b$ is the inverse of the (a)dS
radius, and the $\pm$ sign is for adS and dS, respectively. The
constant $\omega_d$ is equal to $16 \pi/(d-2)\Omega_{d-2}$, where
$\Omega_{d-2}$ is the volume of the unit $d-2$ sphere. The Hawking temperature
of the black hole horizon $r_h$ is
\beq
T_{S(a)dS}= \frac{d-3}{4\pi} \left(\frac{1}{r_h} \pm \gamma^2 r_h\right)
\hbar c\,,
\label{completeADSTemp}
\eeq
where $\gamma$ is proportional to the inverse of the curvature radius of the
(a)dS spacetime
\beq
\gamma=b^{-1}\sqrt{(d-1)/(d-3)}\,.
\eeq
Two limits of the temperature may be realized. In the Schwarzschild limit, the
radius of the event horizon is negligible in comparison to the radius of
curvature of the (a)dS spacetime. The Schwarzschild-(a)dS solution reduces to
the asymptotically Schwarzschild solution with temperature  Eq.~(\ref{TS}). In
the (a)dS limit, the radius of the black hole event horizon is large in
comparison to the radius of curvature of the (a)dS spacetime. The temperature
of the (cosmological) horizon is
\beq
T_{(a)dS}=\frac{(d-3)\gamma^2  r_h}{4\pi}\hbar c\,.
\label{TC}
\eeq
Clearly, the Heisenberg uncertainty principle cannot reproduce Eq.~(\ref{TC}).
However, the (a)dS temperature may be obtained by substituting the standard
Heisenberg relation with its generalized version.

\section{Generalized uncertainty principle}
The generalized version of the Heisenberg uncertainty principle is usually
given by
\beq
\delx \delp \agt \hbar \left[ 1 + \alpha^2 \ell_p^2 \frac{\delp^2}{\hbar^2}
\right]\,,
\label{gup1}
\eeq
where $\ell_p=(\hbar G_d/c^3)^{1/(d-2)}$ is the Planck length, and
$\alpha$ is a numerical constant \cite{Adler:2001vs,gup-bh}.  Equation
(\ref{gup1}) is quite generic, and describes the quantum mechanical uncertainty
when the microscopic structure of spacetime is taken into account.
Non-commutative quantum mechanics \cite{gup-others} and black hole
gedanken-experiments \cite{gup-gedanken} provide heuristic proofs of the
generalized uncertainty principle. The two limits of Eq.~(\ref{gup1}) (see
below) have been derived in the context of string theory in
Refs.~\cite{gup-strings, Konishi:1989wk}.\\

The generalized uncertainty principle (\ref{gup1}) has both low-energy (quantum
mechanical) and high-energy (quantum gravity) limits. The quantum mechanical
limit is obtained when the second term in the r.h.s.~of Eq.~(\ref{gup1}) is
negligible:
\beq
\alpha^2 \ell_p^2 \frac{\delp^2}{\hbar^2} \ll 1\quad
\rightarrow \quad \frac{\delp}{M_p c} \ll \frac{1}{\alpha} \,.
\label{hlim1}
\eeq
where $M_p=[\hbar^{d-3}/(c^{d-5}G_d)]^{1/(d-2)}$ is the Planck mass.
From this limit, it follows that $\alpha=\mathcal{O}(1)$. The quantum gravity
limit is obtained when
\beq
\alpha^2 \ell_p^2 \frac{\delp^2}{\hbar^2} \sim 1 \quad \rightarrow 
\quad \frac{\delp}{M_p c} \sim \frac{1}{\alpha}\,,
\label{glim1}
\eeq
Equation (\ref{gup1}) implies the existence of a minimum length $l_{\rm min}$ of
order of the Planck length. This can be seen by inverting Eq.~(\ref{gup1}):
\beq
\frac{\delx}{2 \alpha^2 \ell_p^2} \left[ 1 - \sqrt{1 - \frac{4 \alpha^2
\ell_p^2}{\delx^2}}\right]  \alt
\frac{\delp}{\hbar} \alt \frac{\delx}{2 \alpha^2 \ell_p^2} \left[ 1 + \sqrt{1 - \frac{4 \alpha^2 \ell_p^2}{\delx^2}}\right] \,
\label{invertgup1}.
\eeq
The lower limit on the uncertainty in position is
\beq
\delx \agt 2 \alpha \ell_p\equiv l_{\rm min}\,.
\eeq
The standard Heisenberg uncertainty relation is obtained when $l_{\rm min}$ is
negligible compared to the scale of the process, i.e.\ when $\Delta x\gg\ell_p$
or $\alpha \rightarrow 0$. In the opposite limit, i.e.\ $\Delta x\sim
l_{\rm min}$, the uncertainty principle reads
\beq
\frac{\Delta p}{M_pc}\sim \frac{\Delta x}{2 \alpha^2\ell_p}\,.
\label{gup2}
\eeq
Equation (\ref{gup2}) holds when strong quantum gravitational effects are
present, and can be derived directly from the conformal invariance property of
the fundamental string \cite{gup-strings, Konishi:1989wk}. In the stringy
regime, the position uncertainty is proportional to the momentum uncertainty.
Equation (\ref{gup1}) is obtained by interpolating Eq.~(\ref{gup2}) with the
standard uncertainty principle. 

Equation (\ref{gup1}) is not the most general form of the generalized
uncertainty principle \cite{Kempf:1993bq}. The symmetry of the symplectic space
suggests to write
\beq
\delx \delp \ge \hbar \left[ 1 + \beta^2  \frac{\delx^2}{\ell_p^2} \right]\,.
\label{gup6}
\eeq
where $\beta$ is a constant parameter. Combining Eq.~(\ref{gup1}) and
Eq.~(\ref{gup6}) we find the general form
\beq
\Delta x\Delta p\gtrsim \hbar\left[1+\alpha^2\ell_p^2
\frac{(\Delta p)^2}{\hbar^2}+
\beta^2\frac{(\Delta x)^2}{\ell_p^2}\right]\,.
\label{gup5}
\eeq
Equation (\ref{gup5}) possesses identical quantum mechanical limit and quantum
gravity limit of Eq.~(\ref{gup1}). Thus Eq.~(\ref{gup5}) is consistent with the
string theory derivation of the generalized uncertainty principle. Derivation
of Eq.~(\ref{gup5}) in non-commutative quantum mechanics is discussed in
Refs.~\cite{Kempf:1993bq}.\\

It is worthwile to discuss in detail the ``dual'' form (\ref{gup6}) of the
generalized uncertainty principle (\ref{gup1}). This will make clear why the
general form of the uncertainty principle, Eq.~(\ref{gup5}), has been mostly
overlooked in the literature in favor of Eq.~(\ref{gup1}). Equation (\ref{gup6}) gives a different interpolation between the quantum
mechanical limit and the quantum gravity limit than
Eq.~(\ref{gup1}). The quantum mechanical limit is obtained when
\beq
\beta \frac{\delx}{\ell_p} \ll 1\quad \rightarrow \quad
\delx \ll \frac{\ell_p}{\beta}
\,.
\eeq
Therefore, it follows that $\beta\ll 1$. The quantum gravity limit is obtained
when
\beq
\beta \frac{\delx}{\ell_p} \approx 1 \quad \rightarrow \quad
\delx \approx \frac{\ell_p}{\beta} \,.
\label{gup9}
\eeq
Since $\beta\ll 1$, one obtains the interesting result that quantum
gravitational effects manifest themselves at very large distances. When the
generalized uncertainty principle was first derived, the idea of modifications
of gravity at great distances had not yet been seriously considered in the
literature. Thus the interpolation (\ref{gup6}) was overlooked. Inverting
Eq.~(\ref{gup6}), 
\beq
\frac{\delp}{2 \beta^2 M_{p} c} \left[ 1 - \sqrt{1 - \frac{4 \beta^2 M_p^2 c^2}
{\delp^2}} \right] \le \frac{\delx}{\ell_p} \le
\frac{\delp}{2 \beta^2 M_{p} c} \left[ 1 + \sqrt{1 - \frac{4 \beta^2 M_p^2
c^2}{\delp^2}} \right] \, ,
\eeq
one obtains a lower bound on the momentum uncertainty. This defines the minimum
momentum $P_{\rm min} = 2 \beta M_p c$.

\section{Schwarzschild-adS thermodynamics with the generalized uncertainty
principle}
The Hawking temperature of the adS black hole can be obtained by repeating the
derivation of Sect.~I with the generalized uncertainty principle. For
semiclassical black holes, $\delx\gg\ell_p$ and $\delp\ll M_{p}c$, and the
form (\ref{gup6}) of the generalized uncertainty principle applies.
If we identify the parameter $\beta$ with $\gamma\ell_p$, Eq.~(\ref{gup6})
reproduces the Schwarzschild-adS Hawking temperature 
\beq
T_{adS} \sim c\Delta {p} \sim \left(\frac{1}{\delx}+\frac{\beta^2}{\ell_p^2}
\delx\right)\hbar c\qquad\to\qquad T_{adS}= 
\frac{d-3}{4\pi}\left(\frac{1}{r_h} + \gamma^2 r_h\right)\hbar c\,.
\label{TADS}
\eeq
The two thermodynamical limits of the Schwarzschild-adS black hole follow from
the two limiting relations between position and momentum ($\Delta p \sim
\hbar/\Delta x$ and $\Delta p\sim \hbar\Delta x/\ell_p^2$) of the generalized
uncertainty principle.\\

The above identification suggests that the Hawking temperature in adS and
Schwarzschild spacetimes may have different origins. Since the adS temperature
can be derived from the high-energy limit of the generalized uncertainty
principle, the adS thermodynamics seems to have a quantum gravitational nature.
It is interesting to note that the generalized uncertainty principle is a
consequence of string theory, which can be consistently formulated in adS
spacetime, whereas there is no consistent formulation of string theory in the
Schwarzschild geometry, where the ordinary uncertainty principle suffices to
derive the black hole thermodynamics.\\

A word of explanation is required on the identification of the inverse adS
radius with the generalized uncertainty principle parameter. In the context of
known generalized uncertainty models, the coefficient of the correction term in
the generalized uncertainty principle is proportional either to the fundamental
gravitational length or the inverse string tension. Whereas the functional form
of the generalized uncertainty principle seems to be rather generic and
model-independent, the exact value of the correction depends on the quantum
gravity states of the specific geometry. In the stringy derivation of
Ref.~\cite{Konishi:1989wk}, for instance, the parameter $\alpha$ of
Eq.~(\ref{gup1}) is inversely proportional to the total momentum uncertainty of
the string in a flat background. If the string propagates in a curved
background, we expect its momentum uncertainty, and thus $\alpha$, to be
different.\\

The Schwarzschild-adS geometry is characterized by two length scales (the
fundamental Planck length and the adS radius). The existence of the latter
allows to set $\beta\propto \ell_p/b$ and $\alpha\propto b/\ell_p$. The exact
proportionality constants can be obtained by matching the quantum gravity
limits of the generalized uncertainty principle to the black hole temperature
in the adS regime, Eq.~(\ref{TADS}). Since $\beta\ll 1$, the first
identification applies to the $b\gg\ell_p$ regime, whereas the second
identification applies to the $b\sim\ell_p$ regime. If the adS quantum states
were known, the exact constant of proportionality between $\alpha$, $\beta$ and
$b$ could be formally derived. Unfortunately, in absence of a definite quantum
gravity theory, the derivation of the exact geometry-dependent generalized
uncertainty principle remains an open issue. A heuristic argument that
illustrates the connection between the generalized uncertainty principle
parameter in the adS spacetime and the adS radius is the following. Let us
suppose to measure the momentum of particle by a scattering with a photon. The
uncertainty in the measurement of the particle momentum is bounded from below
by the value of the photon momentum, $\Delta p \gtrsim p_\gamma\sim
h/\lambda$. Since the photon wavelength cannot exceed the radius of the
spacetime, the minimum uncertainty is $\Delta p \sim h/b$.\\

The generalized uncertainty principle derivation applies also to the
three-dimensional Ba\~nados-Teitelboim-Zanelli (BTZ) black hole
\cite{Banados:wn}
\beq
ds^2 = - \left(-\frac{8G_3M}{c^2}+\frac{r^2}{b^2}\right)c^2\dd t^2 +\left(
-\frac{8G_3M}{c^2}+\frac{r^2}{b^2}\right)^{-1} \dd r^2 + r^2 \dd \phi^2 \,,
\eeq
where the black hole radius is $r_{BTZ}=2b\sqrt{2G_3M}/c$. From Eq.~(\ref{gup2}) we
obtain the Hawking temperature of the BTZ black hole 
\beq
  T_{BTZ} \sim \frac{r_{BTZ}}{2\alpha^2\ell_p}M_pc^2\,,\qquad\to\qquad
  T_{BTZ}= \frac{r_{BTZ}}{2\pi b^2}\hbar c\,,
\label{TBTZ}
\eeq
where $\alpha=b\sqrt{\pi}/\ell_p$.

\section{Schwarzschild-dS thermodynamics with the generalized uncertainty
principle}
The Hawking temperature of the Schwarzschild-dS black hole can be obtained from
Eq.~(\ref{gup6}) by analytical continuation of the parameter $\beta$ into the
imaginary plane. This can be shown to be consistent with the topological
structure of the dS spacetime as follows. For the dS spacetime, the analytic
continuation of Eq.~(\ref{gup6}) reads:
\beq
\delx \delp \agt \hbar \left[ 1 - \beta^2  \frac{\delx^2}{\ell_p^2} \right]\,.
\label{gup7}
\eeq
The inverse of Eq.~(\ref{gup7}) is
\beq
\frac{\delx}{\ell_p} \agt \frac{\delp}{2 \beta^2 M_p c}
\left[ \sqrt{1 + \frac{4 \beta^2 M_p^2 c^2}{\delp^2}} -1 \right]
\label{gup8}
\eeq
Since $\Delta p$ is a positive-definite quantity the position uncertainty is
limited from above by
\beq
\Delta x \alt b\sqrt{\frac{d-3}{d-1}}\,.
\eeq
This relation is a statement that the uncertainty in the measurement of
position may not exceed the size of the de-Sitter spacetime.\\
\section{Conclusions}
We have shown that the uncertainty principle derivation of the Hawking
temperature can be extended to (a)dS-like black holes, provided that we
consider the generalized uncertainty principle instead of the standard
Heisenberg relation. The two thermodynamical limits of Schwarzschild-(a)dS
follow from the quantum-regime limit and the standard limit of the generalized
uncertainty principle. This result seems to indicate different origins for
the thermodynamics of Schwarzschild- and (a)dS-like black holes. This could
explain why only (a)dS-like black holes seem to admit a holographic
description in terms of a dual quantum conformal field theory.

\section*{Acknowledgments}

We are grateful to M.~Cadoni, S.~Hossenfelder and L.~Parker for interesting
discussions.

\end{document}